\documentclass{article}

\usepackage{amsfonts}
\usepackage[margin=1in]{geometry}
\usepackage[affil-it]{authblk}
\usepackage{graphicx}
\usepackage{epstopdf}
\usepackage{cite}
\usepackage[square, numbers, comma, sort&compress]{natbib}
\usepackage{amsmath}
\usepackage{xcolor}

\newcommand{\me}{\mathrm{e}}

\newcommand{\DE}{\mathrm{d}}
\newcommand{\IM}{\mathrm{i}}

\newcommand{\pder}[2]{\frac{\partial#1}{\partial#2}}
\newcommand{\ppder}[2]{\frac{\partial^2#1}{{\partial#2}^2}}
\newcommand{\pmder}[3]{\frac{\partial^2#1}{\partial#2\partial#3}}
\newcommand{\Res}[2]{\mathrm{Res}\left\{#1;#2\right\}}
\newcommand{\Laplinv}[1]{\mathcal{L}^{-1}\left\{#1\right\}}
\newcommand{\Avg}[1]{\left<#1\right>}

\newcommand{\VV}{\mathbf{v}}

\newcommand{\dinteg}[4]{\int_{#1}^{#2}\!{#3}\,\DE #4}

\newcommand{\E}{\varepsilon}

\begin{document}


\title{An asymptotic model for the deformation of a transversely isotropic, transversely homogeneous biphasic cartilage layer}

\author{Gennaro Vitucci%
  \thanks{Corresponding author: \texttt{gev4@aber.ac.uk}}}
\author{Ivan Argatov}
\author{Gennady Mishuris}
\affil{Department of Mathematics, IMPACS, Aberystwyth University, Ceredigion, SY23~3BZ, UK}

\date{}
\maketitle
\section*{Abstract}
In the present paper, an asymptotic model is constructed for the short-time deformation of an articular cartilage layer modeled as transversely isotropic, transversely homogeneous (TITH) biphasic material. It is assumed that the layer thickness is relatively small compared with the characteristic size of the normal surface load applied to the upper surface of the cartilage layer, while the bottom surface is assumed to be firmly attached to a rigid impermeable substrate. In view of applications to articular contact problems it is assumed that the interstitial fluid is not allowed to escape through the articular surface.


\section{Introduction}
\label{sec:intro}
Articular cartilage is a thin tissue which covers the diathrodial joints of the bones. Its structural functions facilitate the transmission of forces between the bones and minimize the stresses contact peaks as well as minimize the friction by means of self-pressurized lubrication. A great interest surrounds its understanding because a correct modeling may lead to correct patient-specific diagnosis for degeneration pathologies and provide operative tools for repair and replacement engineering (see \cite{ateshian2015toward}). A cartilage layer itself is a complex arrangement of a solid matrix saturated by interstitial fluid, mainly composed of water and mobile ions. Its collagen fibrils and proteoglycans are considered the cartilage most relevant solid elements and are heterogeneously distributed along the depth from the subchondral bone to the contact surface. This complex architecture supplies the anisotropic and inhomogeneous electro-mechanical features of the thin structure and is the cause for the nonlinear response to external stimuli.

One approach to the analysis consists in considering the solid phase as a fibril-reinforced material and modeling the full complex layer through a finite element analysis (e.g. \cite{li1999nonlinear, korhonen2003fibril, wilson2005fibril}). A big concern related to the use of the latter arises, concerning contact problems, in modeling thin layers as interphases between structures whose sizes exceed the layers ones of at least one order of magnitude. An extremely fine mesh is required for both the thin layer and the neighbor bone regions, which can easily give place to ill-conditioning and numerical instability of the method if not simply to an enormous increase of the computational effort (see e.g. \cite{wilson2005role, day1994zero, capdeville2008shallow} and relative references). Homogenization procedures are then required in order to provide mathematically workable mechanical laws and they are often obtained, following a long tradition, via multi-scale approaches. Not only the pleasant circumstance of the 70th anniversary, but his great contribution in the field encourage us to mention at least a few works of Professor Federico J.~Sabina in the realm of fiber-reinforced with transversely isotropic constituents \cite{rodriguez2001closed, bravo2001closed, guinovart2001closed, guinovart2005recursive,sabina2001closed,berger2006unit,guinovart2005closed} or laminated \cite{bravo2008homogenization,camacho2009magnetoelectric} materials subjected to elastic, thermal, electrical, magnetic multi-physics. With a homogenized constitutive law in hand, analytical methods to tackle the mechanics of thin layers have been developed. Mainly, they consist in reducing the problem to boundary value problems, thus allowing for the substitution of the finite thickness layer with a zero-thickness one \cite{bovik1994modelling, movchan1995mathematical, klarbring1998asymptotic, mishuris2004imperfect, benveniste2006general, sussmann2011combined} and sometimes even used to improve and make more efficient experimental data extrapolations (e.g. \cite{ochsner2007new,argatov2014small}). Their applicability must be examined case by case since inaccurate assumptions may even lead to non-uniqueness of the solutions which does not derive from the original mathematical ansatz, as proved by Dalla Riva and Mishuris in \cite{dalla2015existence}, however these analytical models can be eventually suitable to be thereafter implemented for asymptotic finite-element computation. A very recent work has been published by Cerfontaine et al. \cite{cerfontaine20153d} about the construction of a zero-thickness homogeneous element which includes the hydro-mechanical coupling.

The debate on the appropriate constitutive model for articular cartilage, when assumed as a continuum medium, is wide, but applications basically count two families. The cartilage material can be considered monophasic and thus its observed delayed response requires a viscoelastic constitutive law \cite{parsons1977viscoelastic, armstrong1986analysis}, or its phenomenology derives from flow-dependent viscoelasticity. The former finds application, for instance, in dynamic \cite{simon1984creep,argatov2013accounting} and impact \cite{GarciaAltieroHaut1998,argatov2015impact} problems for articular cartilage, the latter leads to the development of a biphasic tissue model within the settings of the mixture theory \cite{mow1980biphasic}. It is noticeable that, in terms of response for the impacting body, the two models can be mathematically connected and give nearly the same results \cite{argatov2013mathematical}. The present work is inscribed within the second framework described above. This approach is for instance particularly suited for underlining that the fluid, about 80\% of the structure volume, is the main responsible, for load-bearing at early time of deformation and allows to distinguish between the stresses of the solid structure and the pressure of the interstitial fluid. 
With the purpose of studying the contact problem for the diathrodial joint, analytical solutions for biphasic isotropic homogeneous \cite{ateshian1994asymptotic,wu1997improved,argatov2010axisymmetric,argatov2011elliptical,argatov2011contact,quinonez2011analytical}, elastic and viscoelastic \cite{Barber1990,Eberhardt1990joint,perez2008modified,Lin2010surrogate,argatov2011frictionless,argatov2012development} and transversely isotropic models \cite{rahman2001type,argatov2014small} have been retrieved. Nevertheless, it has been shown that a depth-dependent variation of the solid matrix stiffness and permeability may play a crucial role in determining the internal behavior of the layer. For instance it affects the homogeneity of the stress fields and improves the superficial fluid support in contact solicitation (see \cite{schinagl1997depth, krishnan2003inhomogeneous, federico2005transversely, federico2008anisotropy, ateshian2009modeling, chegini2010time}).

A recent interest developed in mechanics, which involves the study of inhomogeneous structures in the second half of the last century for aerospace or geomechanical purposes. The main reason is to be addressed is the necessity of individuating the response features of composite materials, eventually functionally-graded. A number of analytical studies of inhomogeneous structures have been provided for special material variation functions and for arbitrary inhomogeneity in axisymmetric configuration for monophasic layers. An extended bibliography was examined by Tokovyy and Ma in \cite{tokovyy2015analytical}. To the best of our knowledge, the present work is the first study which, by means of asymptotic analysis, provides an analytical solution for the deformation problem of a biphasic transversely isotropic transversely homogeneous (TITH) thin layer.

An infinitely extended thin porous solid matrix is considered to be linear elastic, the interstitial fluid is inviscid, and the problem is stated within the framework developed by Athesian et al. in \cite{ateshian1994asymptotic} for an isotropic homogeneous layer and by Argatov and Mishuris in \cite{argatovcontact,argatov2015asymptotic} for the transversely isotropic case. The fluid flow is impeded through both the surfaces and the structure is supposed to be firmly attached to a rigid substrate, thus neglecting the influence of the deformability of the substrate, for which an approach was proposed in \cite{argatov2014small}. An arbitrary load is applied to the external surface in absence of friction. Whereas the formulation remains completely general, a special in-depth exponential variation of the stiffness and permeability is assumed. The leading terms of the Laplace transform of  the displacement field and the fluid pressure are retrieved. The boundary conditions used in the deformation problem are thought to be applied to contact problems for which analytical results have been already provided in \cite{argatovcontact}. In this context, explicit formulae are given for the dependent variables only along the external surface, which is of the main interest in contact problem. Numerical benchmarks are studied and compared to the above mentioned existing solutions.

\section{TITH Solid Matrix and Biphasic Model}
\label{sec:model}
The cartilage layer is modelled as a transversely isotropic, transversely homogeneous (TITH) porous linear elastic solid matrix, saturated by a fluid with zero-viscosity. 

The particular boundary condition problem investigated here describes a thin layer completely constrained at the bottom by a flat impermeable surface. A vertical load is applied at its top by a rigid impermeable punch. No friction arises between the punch and the layer upper surface.

While $(x',y')$ are the in-plane coordinates, $z'$, the vertical one, is directed downward, set to 0 at the top.
The solid matrix equilibrium and constitutive equations, coupled with Terzaghi's principle are written as follows:
\begin{gather}
\begin{split}
\pder{}{z'}\left(A_{44}\pder{\VV'}{z'}\right)+A_{13}\pder{\nabla_{y'} w'}{z'}+\pder{}{z'}\left(A_{44}\nabla_{y'} w'\right)+A_{66}\Delta_{y'}\VV'\qquad\qquad\qquad\\+(A_{11}-2A_{66}-A_{12})\nabla_{y'}\nabla_{y'}\cdot\VV+(A_{66}+A_{12})\mathcal{H}_{y'}\VV'=\nabla_{y'}p ,\end{split}\label{eq:equi1}
\\
\pder{}{z'}\left(A_{33}\pder{w'}{z'}\right)+\pder{}{z'}\left(A_{13}\nabla_{y'}\cdot\VV'\right)+A_{44}\pder{\nabla_{y'}\cdot\VV'}{z'}+A_{44}\Delta_{y'}w'=\pder{p}{z'}.\label{eq:equi2}
\end{gather}
In this notation $\mathcal{H}_y$ indicates the Hessian matrix operator whose $ij$-components are $\pmder{}{x'_i}{x'_j}$.

The continuity equation for the fluid and Darcy's law are collected as
\begin{equation}
\pder{}{z'}\left(K_3\pder{p}{z'}\right)-\pder{\nabla_{y'}\cdot\VV'}{t'}-\pmder{w'}{t'}{z'}+K_1\Delta_{y'}p=0.
\label{eq:darc}
\end{equation}
If the thickness of the layer is $h$, the constraint at the bottom surface leads to the boundary conditions
\begin{gather}
\VV'|_{z'=h}=0,\\
w'|_{z'=h}=0,
\label{eq:BCorigd}
\end{gather}
while, the impermeability of both of the bottom and upper surfaces implies
\begin{gather}
\left.\pder{p}{z'}\right|_{z'=h}=0,\\
\left.\pder{p}{z'}\right|_{z'=0}=0.
\end{gather}
The frictionless contact between the rigid punch and the top of the layer allows to state that
\begin{equation}
\left.\pder{\VV'}{z'}+\nabla_{y'}w'\right|_{z'=0}=\mathbf{0}.
\end{equation}
The top surface itself must be also in equilibrium and respect Terzaghi's principle, that is
\begin{equation}
\left.A_{13}\nabla_{y'}\cdot\VV'+A_{33}\pder{w'}{z'}-p\right|_{z'=0}=-q.
\end{equation}
As for the initial conditions, every variable is set to 0 at $t'=0$.


\section{Asymptotic Analysis}

The thinness of the layer suggests to make use of perturbation analysis to solve the system of second order partial differential equations described in Section~\ref{sec:model}.
The thickness $h$ is assumed to be represented as
\begin{equation}
h=\E h_*,
\end{equation}
where $\E$ is a small positive parameter and $h_*$ is a length independent of $\E$ with the the same order of magnitude as the characteristic in-plane length of the loaded layer. 
Thus, it becomes useful: 
\begin{itemize}	 
\item  to introduce the new independent variables
\begin{equation}
z=\frac{z'}{h},\quad t=\frac{t'}{h^2},\quad x_i=\frac{x'_i}{h_*}\quad(i=1,2),
\label{eq:nindv}
\end{equation}
so that $z\in[0,1]$;
\item to set the new unknowns variables
\begin{equation}
w=\frac{w'}{h},\quad \VV=\frac{\VV'}{h};
\label{eq:nunkv}
\end{equation}
\item to express the elastic parameters $A_{jk}$ and the hydraulic resistivities $K_j$ ($j,k=1,2,3$) as functions of the new stretched vertical coordinate $z=\frac{z'}{h}=\frac{z'}{\E h_*}$.
\end{itemize}
The asymptotic expansion of the unknowns is written as follows:
\begin{equation}
\begin{aligned}
\VV&=\E^0 \VV_0+\E^1 \VV_1+\E^2 \VV_2...,\\
w&=\E^0 w_0+\E^1 w_1+\E^2 w_2...,\\
p&=\E^0 p_0+\E^1 p_1+\E^2 p_2...\,.
\end{aligned}
\label{eq:asyexp}
\end{equation}
Substituting \eqref{eq:nindv} and \eqref{eq:nunkv} into Eqs.\eqref{eq:equi1}--\eqref{eq:darc} leads to a new set of differential equations governing the problem
\begin{align}
\begin{split}
\pder{}{z}\left(A_{44}\pder{\VV}{z}\right)+\E\left(A_{13}\pder{\nabla_{y} w}{z}+\pder{}{z}\left(A_{44}\nabla_{y} w\right)-\nabla_{y}p\right)\qquad\qquad\\+\E^2\left( (A_{11}-2A_{66}-A_{12})\nabla_{y}\nabla_{y}\cdot\VV+(A_{66}+A_{12})\mathbf{H}_{y}\VV\right)=\mathbf{0},\end{split} \label{eq:nasy1}
\\
&\pder{}{z}\left(A_{33}\pder{w}{z}\right)-\pder{p}{z}+\E\left(\pder{}{z}\left(A_{13}\nabla_{y}\cdot\VV\right)+A_{44}\pder{\nabla_{y}\cdot\VV}{z}\right)+\E^2A_{44}\Delta_{y}w=0,
\label{eq:nasy2}\\
&\pder{}{z}\left(K_3\pder{p}{z}\right)-\pmder{w}{t}{z}+\E\left(-\pder{\nabla_{y}\cdot\VV}{t}\right)+\E^2K_1\Delta_{y}p=0.\label{eq:nasy3}
\end{align}
In the same way, the boundary conditions take the form 
\begin{align}
\VV|_{z=1}=\mathbf{0}, \quad w|_{z=1}=0,\\
\left.\pder{p}{z}\right|_{z=1}=0, \quad
\left.\pder{p}{z}\right|_{z=0}=0,\\
\left.\pder{\VV}{z}+\E\nabla_{y}w\right|_{z=0}=\mathbf{0},\\
A_{33}\pder{w}{z}-p+q+\E\left.A_{13}\nabla_{y}\cdot\VV\right|_{z=0}=0. \label{eq:bc4asy}
\end{align}
Using the expansions \eqref{eq:asyexp} to solve the system \eqref{eq:nasy1}--\eqref{eq:nasy3} and taking into account the boundary conditions, it is easy to verify that the trivial terms of the expansions are
\begin{gather}
\VV_0=\VV_2=\mathbf{0},\quad w_0=w_1=p_1=0,\\
p_0=q,
\label{eq:trivterm}
\end{gather}
so that
\begin{equation}
\VV_1=\nabla_{y}q\dinteg{1}{z}{\frac{z}{A_{44}}}{z}.
\label{eq:vv1impl}
\end{equation}
The $\E^2$-terms of Eqs.~\eqref{eq:nasy2} and \eqref{eq:nasy3} yield
\begin{gather}
\pder{}{z}\left(A_{33}\pder{w_2}{z}\right)+\pder{}{z}\left(A_{13}\nabla_{y}\cdot \VV_1\right)-\pder{p_2}{z}+A_{44}\pder{\nabla_{y}\cdot \VV_1}{z}=0 ,\label{eq:n22}\\
\pder{}{z}\left(K_3\pder{p_2}{z}\right)-\pmder{w_2}{t}{z}-\pder{\nabla_{y}\cdot \VV_1}{t}+K_1\Delta_{y}q=0 .\label{eq:n32}
\end{gather}
The boundary condition \eqref{eq:bc4asy} becomes
\begin{equation}
\left.A_{33}\pder{w_2}{z}+A_{13}\nabla_{y}\cdot \VV_1-p_2\right|_{z=0}=0.
\end{equation}
Integrating Eq.~\eqref{eq:n22} once between 0 and $z$ and applying the latter boundary condition leads to the following equation:
\begin{equation}
A_{33}\pder{w_2}{z}=p_2-\Delta_{y}q\frac{z^2}{2}-A_{13}\nabla_{y}\cdot \VV_1
.\label{eq:w2fromp2}
\end{equation}
Thanks to Eq.~\eqref{eq:vv1impl} and the equation above, Eq.~\eqref{eq:n32} can be expressed exclusively in terms of $p_2$ as

\begin{equation}
\begin{split}
\ppder{p_2}{z}-\frac{1}{K_3}\pder{K_3}{z}\pder{p_2}{z}-\frac{1}{K_3 A_{33}}\pder{p_2}{t}=\\
\frac{1}{K_3 A_{33}}\pder{\Delta_y q}{t}\left((A_{33}-A_{13})\dinteg{1}{z}{\frac{z}{A_{44}}}{z}-\frac{z^2}{2}\right)-\frac{K_1}{K_3}\Delta_y q,
\end{split} \label{eq:n2+3}
\end{equation}
where the unknowns are kept on the left-hand side.
\section{Laplace transformation in the case of a specific type of inhomogeneity}
\label{1otSection4}

Some assumption on the variation of the five parameters $A_{13}$, $A_{33}$, $A_{44}$, $K_1$, and $K_1$ needs to be done in order to simplify the continuation. In the present work we consider that they vary exponentially along the $z$-axis while the product $K_3 A_{33}$ remains constant. The latter feature is validated by experimental observation which suggests that the axial mechanical stiffness --- then $A_{33}$ --- increases with the depth from the surface toward the bone \cite{schinagl1997depth,klein2007depth,wang2001analysis}, while the fact that a decreasing porosity causes an overall reduction in permeability was shown in \cite{federico2008anisotropy}, where, through this hypothesis, the classical results of \cite{maroudas1968permeability} were fitted and justified. The exponential depth-dependency used later on is
\begin{equation}\left\{\begin{aligned}
&A_{33}=a_{33}\me^{2\gamma z},\quad A_{44}=a_{44}\me^{\alpha z},\quad A_{13}=a_{13}\me^{\alpha_{13}z},\\
&K_3=k_3\me^{-2\gamma z},\quad K_1=k_1\me^{-\gamma_1 z},
\end{aligned}\right.
\label{eq:paramdef}
\end{equation}
where $\gamma>0$ is a specified constant. Certainly the choice of exponential functions is not completely general, nevertheless, given that we deal with a very thin layer, the opportunity of fitting experimental data does not appear particularly compromised, at least in the case of monotonic variations of the parameters in exam.
The latter expressions are substituted into Eq.~\eqref{eq:n2+3} and the time variable $t$ is changed into the dimensionless one by the formula $$\tau=K_3 A_{33} t.$$ 
Recalling that any unknown is set to 0 for $-\infty<t<0^-$, the inverse Laplace transformation is applied to yield 

\begin{equation}
\begin{split}
\ppder{P}{z}-2\gamma\pder{P}{z}-s P=\\
s\Delta_y Q\left(\frac{a_{33}\me^{2\gamma z}-a_{13}\me^{\alpha_{13}z}}{a_{44}}\dinteg{1}{z}{\frac{z}{\me^{\alpha z}}}{z}-\frac{z^2}{2}\right)-\frac{k_1}{k_3}\me^{(2\gamma-\gamma_1)z}\Delta_y Q,
\end{split} \label{eq:Ln2+3}
\end{equation}
where $P(s)$ and $Q(s)$ are respectively the Laplace transforms of $p_2(\tau)$ and $q(\tau)$,  $s$ is the transformation parameter.
For the terms that multiply an exponential function of $z$ we introduce the following abbreviation:
\begin{equation}
\Phi^{(M)}_i=\me^{M_i z}(b_{i1}z+b_{i0})\quad(i=1,2,3,4),
\end{equation}
whose coefficients are resumed in Table~\ref{tab:phiM}.
\begin{table}
\begin{center}
\begin{tabular}{c}
$\begin{array}{|c||c|c|c|}
\hline
i&M_i&b_{i1}\alpha^2a_{44}&b_{i0}\alpha^2a_{44}\\
\hline\hline
1&\alpha_{13}-\alpha&\alpha a_{13}&a_{13}\\
\hline
2&\alpha_{13}&0&-a_{13}(1+\alpha)\me^{-\alpha}\\
\hline
3&2\gamma-\alpha&-\alpha a_{33}&-a_{33}\\
\hline
4&2\gamma&0&a_{33}(1+\alpha)\me^{-\alpha}\\
\hline
\end{array}$
\end{tabular}
\caption{Parameters of $\Phi^{(M)}_i$}\label{tab:phiM}
\end{center}
\end{table}
It ensues that the latter second order ordinary differential equation is suitable to be rewritten as
\begin{equation}
\ppder{P}{z}-2\gamma\pder{P}{z}-s P=\Delta_y Q \sum_{i=1}^6\Upsilon_i(s,z) .\label{eq:LMn2+3}
\end{equation}
Here we have introduced the notation 
\begin{equation}
\begin{aligned}
&\Upsilon_i(s,z)=s\Phi^{(M)}_i(z),\quad i=1,2,3,4,\quad\\
&\Upsilon_5(s,z)=-s\frac{z^2}{2},\quad
\Upsilon_6(s,z)=-\frac{k_1}{k_3}\me^{(2\gamma-\gamma_1)z}.
\end{aligned}\label{eq:Upsilon}
\end{equation}
Posed $\sigma(s)=\sqrt{\gamma^2+s}$, the homogeneous solution of Eq.~\eqref{eq:LMn2+3} is
\begin{equation}
P_h=\me^{\gamma z} (C_1 \sinh\sigma z+C_2\cosh\sigma z),
\label{eq:HomogP}
\end{equation}
where the two constants $C_1(s)$ and $C_2(s)$ must be determined to fulfill the boundary conditions $\pder{P}{z}=0$ at $z=0$ and $z=1$.

Within the Appendices \ref{1otAppendix_A}, \ref{1otAppendix_B} and \ref{1otAppendix_C}, we deal with the solution of Eq.~\eqref{eq:LMn2+3} splitting it into three parts (see Eq.~\eqref{eq:Upsilon}) as follows: one part containing the so-called $\Phi^{(M)}_i$-terms, corresponding to $\Upsilon_i(s,z)$, $i=1,2,3,4$, one containing the $z^2$-term, corresponding to $\Upsilon_5(s,z)$, and the last one which involves the permeability, specifically the ratio $\frac{K_1(z)}{K_2(z)}$. This term, corresponding to $\Upsilon_6(s,z)$, will take the name of $k$-term.

It is useful to calculate (with the same subdivision) the integral $$\xi=\dinteg{1}{z}{p_2\me^{-2\gamma z}}{z}$$ which comes out from Eq.~\eqref{eq:w2fromp2}, if $w_2$ is recovered and the boundary condition $w_2=0$ at $z=1$ is applied.
\begin{equation}
w_2=\dinteg{1}{z}{\dfrac{\me^{-2\gamma z}\left(p_2-\Delta_{y}q\frac{z^2}{2}-A_{13}\nabla_{y}\cdot \VV_1\right)}{a_{33}}}{z}.
\label{eq:Iw2fromp2}
\end{equation}
In particular we are interested in the evaluation of the pressure and the vertical displacement at the load application surface ($z=0$) because those results are especially important for contact problems.

The asymptotic expansion for the fluid pressure at the load application surface ($z=0$) results from Eq.~\eqref{eq:asyexp} and Eq.~\eqref{eq:trivterm}:
\begin{gather}
p_0\approx q(\tau)+\frac{h^2}{h_\ast^2}p_{02}(\tau),\label{eq:p0tot}\\
p_{02}=\sum_{i=1}^4 p_{0i}^{(M)}+p_0^{(2)}+p_0^{(k)},
\label{eq:p0sum}
\end{gather}
where the terms $p_{0i}^{(M)}$, $p_0^{(2)}$ and $p_0^{(k)}$ must be calculated as derived in the Appendices \ref{1otAppendix_A}, \ref{1otAppendix_B} and \ref{1otAppendix_C}.

Let us assume that both $\alpha_{13}$ and $\gamma$ are zero. In this case Table~\ref{tab:phiM} becomes Table~\ref{tab:phiM00}. For studying the homogeneous case $\alpha$ must also tend to zero, so that
\begin{table}
\begin{center}
\begin{tabular}{c}
$\begin{array}{|c||c|c|c|}
\hline
i&M_i&b_{i1}\alpha^2a_{44}&b_{i0}\alpha^2a_{44}\\
\hline\hline
1&-\alpha&\alpha a_{13}&a_{13}\\
\hline
2&0&0&-a_{13}(1+\alpha)\me^{-\alpha}\\
\hline
3&-\alpha&-\alpha a_{33}&-a_{33}\\
\hline
4&0&0&a_{33}(1+\alpha)\me^{-\alpha}\\
\hline
\end{array}$
\end{tabular}
\caption{Parameters of $\Phi^{(M)}_i$ for $\alpha_{13}=\gamma=0$}\label{tab:phiM00}
\end{center}
\end{table}
\begin{equation}
-\sum_{i=1}^4b_{i0}\Delta_yq=\Delta_yq\frac{a_{33}-a_{13}}{a_{44}}\lim_{\alpha\rightarrow 0}\frac{1-(1+\alpha)\me^{-\alpha}}{\alpha^2}=\frac{1}{2}\frac{a_{33}-a_{13}}{a_{44}}\Delta_yq,
\end{equation}
\begin{equation}
\sum_{i=1}^4\sum_{n=0}^\infty{\mathrm{Res}\left\{\me^{s\tau}\Omega_i^{(M)}(s);s_{n}\right\}}\ast\Delta_yq=\frac{a_{33}-a_{13}}{a_{44}}2\sum_{n=0}^\infty(-1)^n\me^{-n^2\pi^2\tau}\ast\Delta_yq.
\end{equation}
The last two equations imply that the homogeneous solution regarding the $\Phi^{(M)}_i$-terms is written as
\begin{equation}
\sum_{i=1}^4 p_{0i}^{(Mh)}=\frac{1}{2}\frac{a_{33}-a_{13}}{a_{44}}\Delta_yq+\frac{a_{33}-a_{13}}{a_{44}}2\sum_{n=0}^\infty(-1)^n\me^{-n^2\pi^2\tau}\ast\Delta_yq.
\end{equation}
Analyzing the same limits for the part of the solution generated by the $z^2$-term, we get
\begin{equation}
p_0^{(2h)}=-2\sum_{n=0}^\infty(-1)^n\me^{-n^2\pi^2\tau}\ast\Delta_yq,
\end{equation}
and for the $k$-term, we obtain 
\begin{equation}
p_0^{(kh)}=\frac{k_1}{k_3}\ast\Delta_yq.
\end{equation}
Collecting the three formulas above and substituting them into Eqs.~\eqref{eq:p0tot} and \eqref{eq:p0sum}, the following complete expression for the $\varepsilon^2$-approximation of the fluid pressure at $z=0$ can be achieved:
\begin{equation}
p_0^h\approx q(\tau)+\frac{h^2}{h_\ast^2}p_{02}^h(\tau)\label{eq:p0htot},
\end{equation}
\begin{equation}
\begin{split}
p_{02}^h&=\frac{1}{2}\frac{a_{33}-a_{13}}{a_{44}}\Delta_yq(\tau)\\
&+\frac{k_1}{k_3}\dinteg{0}{\tau}{\Delta_yq(\theta)}{\theta}\\
&+2\left(\frac{a_{33}-a_{13}}{a_{44}}-1\right)\sum_{n=0}^\infty(-1)^n\dinteg{0}{\tau}{\me^{-n^2\pi^2(\tau-\theta)}\Delta_yq(\theta)}{\theta}.
\end{split}
\label{eq:p0hsum}
\end{equation}
This is exactly the same expression obtained by Argatov and Mishuris in \cite{argatovcontact}.

Equation~\eqref{eq:Iw2fromp2} shows how to obtain the $\varepsilon^2$-term of the asymptotic expansion of the vertical displacement $w_2$ --- which is actually also the only non-zero term of the approximation of $w$ (see Eq.~\eqref{eq:trivterm}) --- starting from $p_2$ and $\VV_1$. Sticking to the coordinate $z=0$, the mentioned equation can be written as
\begin{equation}
\begin{split}
\frac{w_0}{\varepsilon^2}\approx w_{02}&=\frac{1}{a_{33}}\left(\sum_{i=1}^4\xi_{0i}^{(M)}+\xi_0^{(2)}+\xi_0^{(k)}\right)+\frac{\Delta_yq}{2a_{33}}\dinteg{0}{1}{\me^{-2\gamma z}z^2}{z}\\
&+\frac{a_{13}\Delta_yq}{a_{33}a_{44}}\dinteg{0}{1}{\me^{(\alpha_{13}-2\gamma)z}\left(\dinteg{z}{1}{\me^{-\alpha\tilde{z}}\tilde{z}}{\tilde{z}}\right) }{z}.
\end{split}
\end{equation}
It is easy to notice, making use of Eq.~\eqref{eq:xi0QMt}, Table~\ref{tab:phiM}, Eqs.~\eqref{eq:xi02t} and \eqref{eq:xi0kt}, that the only terms which do not vanish are
\begin{equation}\begin{split}
w_{02}&=\frac{\xi_{03}^{(M)}+\xi_{04}^{(M)}+\xi_0^{(k)}+b_{11}M_1(\me^{M_1-2\gamma}-1)\me^{A_2\tau}\ast\Delta_yq(\tau)}{a_{33}}\\
&=\frac{\me^{-\alpha}(\alpha^2+2\alpha+2)-2}{\alpha^3a_{44}}\Delta_yq\\
&+\frac{a_{13}(\alpha_{13}-\alpha)(\me^{\alpha_{13}-\alpha-2\gamma}-1)}{\alpha a_{33}a_{44}}\Delta_yq\ast\me^{(\alpha_{13}-\alpha)(\alpha_{13}-\alpha-2\gamma)\tau}                     \\
&+(\alpha-2\gamma)\frac{\me^{-\alpha}-1}{\alpha a_{44}}\Delta_yq\ast\me^{\alpha(\alpha-2\gamma)\tau}\\
&+\frac{k_1}{a_{33}k_3}\frac{\me^{-\gamma_1}-1}{\gamma_1}\ast\Delta_yq(\tau).
\end{split}
\end{equation}
When all the exponents defined in Eq.~\eqref{eq:paramdef} are set to zero, the previous equation takes the form 
\begin{equation}
w_{02}= -\frac{1}{3a_{44}}\Delta_yq(\tau)-\frac{k_1}{a_{33}k_3}\dinteg{0}{\tau}{\Delta_yq(\theta)}{\theta}.
\end{equation}
As for Eq.~\eqref{eq:p0hsum}, analysing the behavior of a homogeneous transversally isotropic layer, Argatov and Mishuris \cite{argatovcontact} gained the same result. Recovering all the original variables from Eqs.~\eqref{eq:nindv}, \eqref{eq:nunkv}, and \eqref{eq:asyexp}, and writing the load as $q=q(x',y',t')$, we find 
\begin{equation}\begin{split}
w'_{02}
&=\frac{\me^{-\alpha}(\alpha^2+2\alpha+2)-2}{\alpha^3a_{44}}h^3\Delta_{y'}q\\
&+\frac{a_{13}(\alpha_{13}-\alpha)(\me^{\alpha_{13}-\alpha-2\gamma}-1)}{\alpha a_{44}}h k_3\dinteg{0}{t'}{\me^{(\alpha_{13}-\alpha)(\alpha_{13}-\alpha-2\gamma)\frac{a_{33}k_3}{h^2}(t'-\theta)}\Delta_{y'}q}{\theta}                     \\
&+(\alpha-2\gamma)\frac{\me^{-\alpha}-1}{\alpha a_{44}}h k_3 a_{33}\dinteg{0}{t'}{\me^{\alpha(\alpha-2\gamma)\frac{a_{33}k_3}{h^2}(t'-\theta)}\Delta_{y'}q}{\theta}\\
&+\frac{\me^{-\gamma_1}-1}{\gamma_1}h k_1 \dinteg{0}{t'}{\Delta_{y'}q(\theta)}{\theta}.
\end{split}
\label{eq:displ}
\end{equation}

\section{Numerical examples}

In this Section, we present some numerical examples with the main purpose to underline the effect of the inhomogeneity to the response of the cartilage layer to an applied load. For this reason we compare every benchmark to the results obtained considering the solution by Argatov and Mishuris \cite{argatovcontact} for a transversely isotropic homogeneous (TIH) model with the same average permeability and mechanical stiffness.

The thickness of the layer is taken to be $h=10$mm (for the sake of easiness of scaling). The applied distributed load $q=q_{t'} q_r$ is axisymmetric with respect to the radial coordinate $r$, though this symmetry (not necessary from the assumptions) represents a lack of generality only for the sake of clarity in this section. It results from the product of two factors: $q_{t'}=q_{t'}(t')$ assigns the behavior in time, while $q_r=q_r(x',y')=q_r(r)$ contributes to the spacial distribution.

A total force $F=125$N is distributed according to the law
\begin{equation}
q_r(r)=\me^{-\left(\frac{1.73 r}{10 h}\right)^2}\left(\frac{10 h}{1.73}\right)^2\frac{F}{\pi},
\label{eq:spaF}
\end{equation}
so that 100N are loaded within a radius of about $10h$.

A homogeneous and isotropic Poisson's ratio $\nu=0$ is considered, thus the stiffness parameters (see Eq.~\eqref{eq:paramdef}) result as follows:
\begin{equation}
\left\{\begin{aligned}
A_{33}&=H_{A3}(z')=a_{33}\me^{2\gamma z'/h},\\
A_{13}&=\frac{\nu}{1-\nu}H_{A1}(z')=a_{13}\me^{\alpha_{13}z'/h},\\
A_{44}&=\frac{1-2\nu}{2(1-\nu)}\text H_{A1}(z')=a_{44}\me^{\alpha z'/h},
\end{aligned}\right.
\end{equation}
where $H_{A1}$ and $H_{A3}$ are respectively the planar and the vertical aggregate moduli. For the fixed $\nu$, the behavior of $A_{33}$ and $A_{13}$ must be the same and only due to the variation of $\text{Ha}_1$, so that $\alpha_{13}=\alpha$.

The examples show a cartilage layer which is in average isotropic both in aggregate modulus and permeability. Using the operator $\Avg{\cdot}$ to express average along the depth:
\begin{equation}
\left\{\begin{aligned}
\Avg{H_{A1}}=\Avg{H_{A3}}=\Avg{H_{A}}&=0.5\text{MPa},\\
\Avg{K_1}=\Avg{K_3}=\Avg{K}&=2\cdot 10^{-15}\frac{\text{m}^2}{\text{Pa s}}.
\end{aligned}\right.
\label{eq:setqua}
\end{equation}
The parameter that is used both to describe the inhomogeneity of the permeability $K_3$ to the one of the stiffness $A_{33}$ is $\gamma$. Anyway one can use a more intuitive quantity (called the \textit{ratio of inhomogeneity}) $R_I$, which says how much $A_{33}$ grows from the articular surface to the bone (i.e., $\me^{2\gamma}$) and at the same time the ratio of $K_3$ at $z'=0$ and at $z'=h$. Thus, we put
\begin{equation}
\gamma=0.5\log R_I.
\label{eq:setg}
\end{equation}
In order to study the effects of the inhomogeneity, we set the remaining parameters as functions of the same $R_I$ as follows:
\begin{equation}
\left\{\begin{aligned}
\gamma_1&=2\log R_I,\\
\alpha=\alpha_{13}&=0.7 \log R_I .
\end{aligned}\right.
\label{eq:setexp}
\end{equation}


\begin{figure}
\centering
\includegraphics[scale=.75]{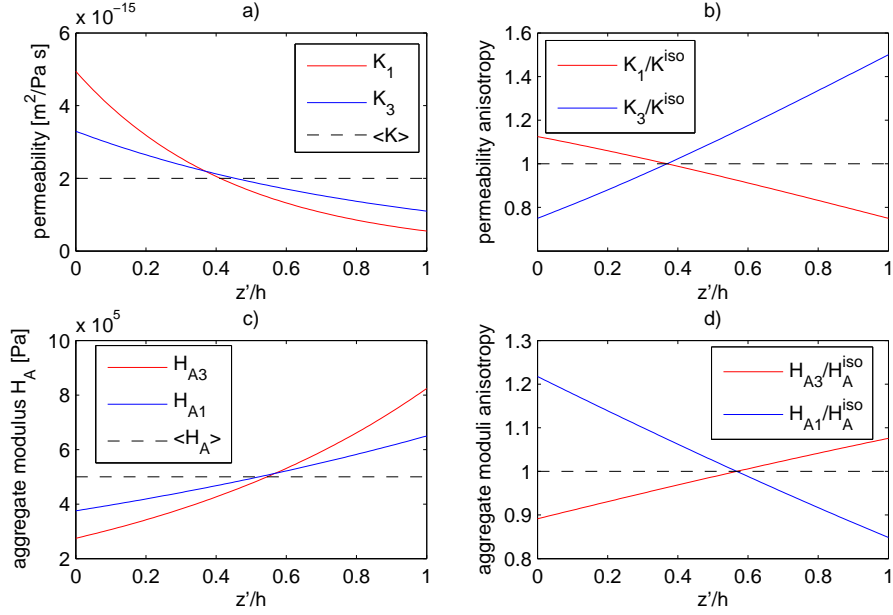}
\caption{Permeability and aggregate modulus versus the isotropic homogeneous ones plotted along the depth $z'/h$ for an \textit{inhomogeneity ratio} $R_I=3$ in a) and c) following the settings of Eq.~\eqref{eq:setqua}, \eqref{eq:setg} and \eqref{eq:setexp}; Plots b) and d) illustrate the descending anisotropy as ratios between those quantities and the equivalent inhomogeneous isotropic ones.}
\label{fig:RA3param}
\end{figure}
As shown by Federico and Herzog in \cite{federico2008anisotropy} via a micromechanical approach, the anisotropy of permeability can be explained through the fact that the collagen fibers, whose statistical orientation varies with the depth, are impermeable. Consequently, being the fibers nearly parallel to the surface and perpendicular to the tidemark, $K_1>K_3$ for small $z'$ and vice versa. Since the same fibers are known to be responsible also for the mechanical properties, and particularly for the anisotropy and inhomogeneity of the cartilage stiffness \cite{federico2005transversely}, it is expected that $H_{A1}>H_{A3}$ in the upper part of the layer, conversely in the lower one. The achievement of these features are the reason of the choice of the parameters above. For instance, the effect of this characterization is visualized in Fig.~\ref{fig:RA3param} for $R_I=3$, where ${K}^{\text{iso}}$ and ${H}^{\text{iso}}_{A}$ (see Fig.~\ref{fig:RA3param}.b and Fig.~\ref{fig:RA3param}.d) are the equivalent inhomogeneous isotropic aggregate modulus and permeability defined as
\begin{equation}
\left\{\begin{aligned}
{K}^{\text{iso}}(z')&=\frac{2}{3}K_1+\frac{1}{3}K_3,\\
{H}^{\text{iso}}_{A}(z')&=\frac{2}{3}{H}_{A1}+\frac{1}{3}{H}_{A3}.
\end{aligned}\right.
\label{eq:}
\end{equation}
\begin{figure}
\centering
\includegraphics[clip=true, keepaspectratio, scale=.70 ]{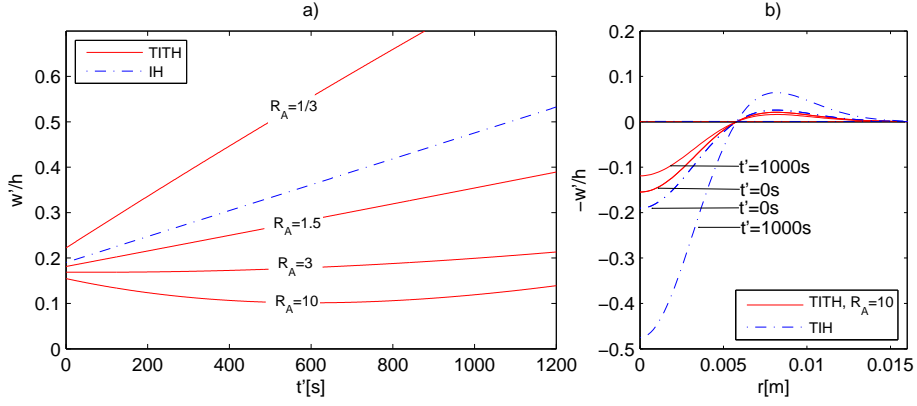}
\caption{Vertical displacement of the TITH layer surface under a constant load and comparison with an IH layer behavior: a) is the evolution in time at $r=0$; b) shows the deformation profile along the radius for different times.}
\label{fig:const1}
\end{figure}
According to Eq.~\eqref{eq:displ}, in the first place we show the behavior of the present cartilage model for a constant load ($q_{t'}=1$) applied for 1200s, smaller than the characteristic time $t'={h^2}/({a_{33} k_3})$ for which the solution is valid. Both from Fig.~\ref{fig:const1}.a and Fig.~\ref{fig:const1}.b it is observable that $R_I$ strongly influences the response of the structure. In particular, while in the case of a homogeneous isotropic layer the deformation increases for a constant load, above a certain value of $R_I$, which depends on the parameters settings, a phenomenon of swelling appears during the initial phase. It derives (see the second and third addends in Eq.~\eqref{eq:displ}) from the contribution of $K_3$, effect that vanishes in the IH case. 
Fig.~\ref{fig:const1}.b draws attention to the profile of deformation of the contact surface. The obtained asymptotic solution provides that $w'$ depends only on $\Delta_{y'}q$, so that, since for our loading condition its zeroes remain fixed (see Eq.~\eqref{eq:spaF}), every benchmark calculated in this section implies a homothetic deformation profile.
\begin{figure}
\centering
\includegraphics[clip=true, width=.9\textwidth]{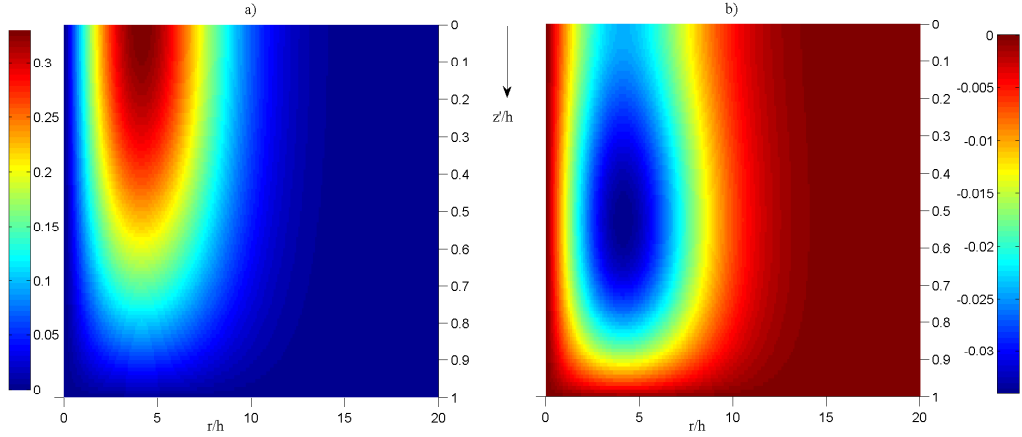}
\caption{a) Lateral displacement of a TITH layer with $R_I=3$ under constant load at $t'=0$; b) Difference of lateral displacement between a TITH layer with $R_I=3$ and an IH one under constant load at $t'=0$.}
\label{fig:constRI3ab}
\end{figure}

Through Eq.~\eqref{eq:vv1impl} the lateral displacements are achieved. In Fig.~\ref{fig:constRI3ab}.a, the axisymmetric $v'$ is plotted under a constant load from $r=0$ to $r=20h$ at $t'=0s$. The highest displacement is obtained at the load surface at $r\approx 4h$ with a value of $v'\approx 0.33 h$ while the base is constrained (see Eq.~\eqref{eq:BCorigd}). In Fig.~\ref{fig:constRI3ab}.b, the difference between the solution for a TITH layer with $R_I=3$ and the one for a IH layer is measured. The maximum difference appears at the same $r\approx 4h$ but $z'\approx 0.5h$,  and is about $0.034h$, so that, in terms of lateral displacements, the TITH structure with $R_I=3$ shows to deform less than the equivalent IH one. In Fig.~\ref{fig:constRI3ab}, the instantaneous response is calculated, and one can notice how (although the surface displacements can appear qualitatively similar at the load surface (Fig.~\ref{fig:const1}.b) and the same fittings are eventually possible both through a homogeneous and inhomogeneous model calibrating the material parameters) remarkable differences are returned between the two if light needs to be shed inside the layer. 

\begin{figure}
\centering
\includegraphics[clip=true, keepaspectratio, scale=.70 ]{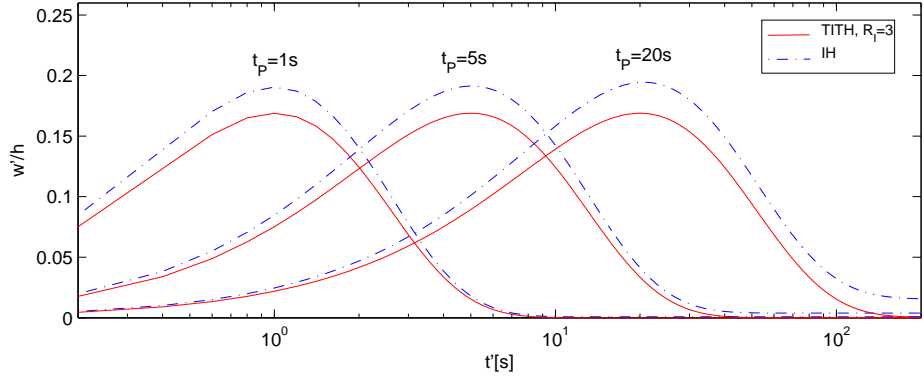}
\caption{Displacement at $r=0$. Peaks of the three considered loads: 1s, 5s, 20s. The representation timescale is logarithmic. The TITH has $R_I=3$ and does not present significant residual displacements at $t'=200s$. Its peak displacements verify at the respective $t_P$ and are about $0.17h$, while the peaks for IH are $0.19h$.}
\label{fig:texpt}
\end{figure}
\begin{figure}
\centering
\includegraphics[clip=true, keepaspectratio, scale=.70 ]{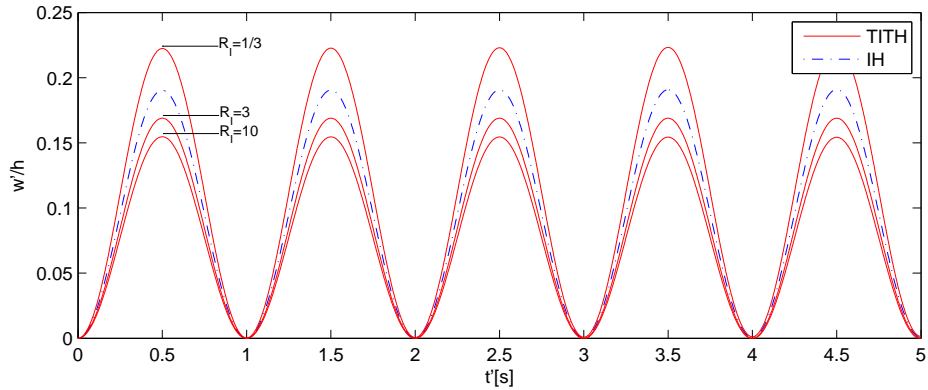}
\caption{Vertical displacement at $r=0$ under a sinusoidal load oscillating with a frequency of 1Hz. Three different inhomogeneous layers ($R_I=1/3,3,10$) are compared to an IH one.}
\label{fig:sine5th}
\end{figure}
The second case that we consider deals with a load which reaches its peak at $t'=t_P$. Successively it decreases to 0 asymptotically following the law:
\begin{equation}
q_{t'}=\frac{t'\me^{-(\frac{t'}{t_P}+1)}}{t_P}.
\label{eq:}
\end{equation}
The displacement of the point $r=0$ is depicted in Fig.~\ref{fig:texpt} for three different values of $t_P$ for the first 200s. The peaks of the displacement are the same in the three cases and happen approximately at the respective $t_P$. As in the case of a constant load, the deformation at $r=0$ becomes smaller for inhomogeneities with $R_I>1$. The difference between the IH model and the TITH ($R_I=3$) model consists mainly in the behavior at large $t'$. While the transversely homogeneous layer returns to the undeformed configuration after the load removal, the homogeneous solution presents a residual deformation that depends on $t_P$, that is on the rate with which the load is applied. 

Finally, in Fig.~\ref{fig:sine5th} the effect on $w'|_{r=0}$ of a sinusoidal load is plotted. The frequency applied is considered to be 1Hz, similar to the one that can occur to a knee articular cartilage due to walking. As expected, since the period of 1s is small in comparison to the characteristic time, the short term difference between differently inhomogeneous layers is exclusively in terms of amplitude and no residual displacements are accumulated for next cycles. The structure deforms as a monophasic elastic one; yet again a positive $R_I$ produces a \textit{stiffer} response, while a negative one vice versa.

\section{Discussion and conclusions}

An analytical approach is provided for solution of the deformation problem of a TITH biphasic thin layer. The mathematical analysis is conducted by use of Laplace transformation and asymptotic analysis. The leading terms of the displacement and fluid pressure fields are retrieved through the solution of ordinary differential equations. Such equations are made particularly simple thanks to the assumption of exponential in-depth variation of the solid matrix elastic stiffness and permeability with the only restriction of keeping the product $k_3 A_{33}$ constant along the layer transverse direction. This particular setting appears reasonable since experimental investigations on articular cartilage show that the aggregate modulus, contrarily to the permeability, decreases toward the subchondral bone (see \cite{wang2001analysis, maroudas1968permeability}). The scope of the present work is presenting an explicit form for the deformation of the external cartilage surface which can be straightforwardly applied for solving contact problems. It is reached through the formula in Eq.~\eqref{eq:displ}. In addition to the contribution of $A_{33}$, $A_{44}$, $k_1$ and $k_3$ found by Argatov and Mishuris in \cite{argatovcontact}, the one of $A_{13}$ is shown, other than the effects of the variation parameters $\alpha_{ij}$ and $\gamma_i$ (see their definitions in Eq.~\eqref{eq:paramdef}). 
As discussed in Section~\ref{sec:intro}, the role of inhomogeneity and anisotropy in affecting the internal state of the cartilage layer during loading encouraged many authors to develop fully 3D models for its mechanical analysis. However their applicability for the study of contact problems, due to the large difference in scales between the thin tissue and the bones interacting along the articular joint may result arduously suitable because of the deriving numerical problems, possibly obligating to use homogeneous elements as interphases.
The simplicity of Eq.~\eqref{eq:displ}, on the opposite, suggests that the same constitutive equations can be used both for an insight into the layer (see Fig.~\ref{fig:constRI3ab}), for instance for experimental investigations, and for a large scale contact problems. The full-thickness layer can finally be substituted by a zero-thickness one through transmission conditions. Eventually an asymptotic-based finite element can be implemented for assessing patient-specific problems for real diathrodial joints and complex geometries, once the material parameters are experimentally estimated. Only in sight of a future application to contact problems the results are presented extensively on the contact surface, while, for the reader who desired to obtain the full-depth solution, it would be enough to remove the restriction on the $z$-coordinate for the Laplace inversion shown in the Appendices.

\section*{Acknowledgments}

GV participated to the present work under the support of \textit{FP7-MC-ITN-2013-606878-CERMAT2}, IA and GM acknowledge \textit{H2020-MSCA-RISE-2014-644175-MATRIXASSAY}.
\bibliographystyle{abbrv}
\bibliography{Bibliography}


\appendix
\section{Solution to Eq.~\eqref{eq:LMn2+3} for the right-hand sides $\Upsilon_i(s,z)$, $i=1,2,3,4$}
\label{1otAppendix_A}
Let us consider the equation
\begin{equation}
\ppder{P}{z}-2\gamma\pder{P}{z}-s P=
s\Delta_y Q\me^{M z}(b_{1}z+b_{0}),
\label{eq:Mp2}
\end{equation}
with the boundary conditions $\pder{P}{z}=0$ at $z=0$ and $z=1$.
Its general solution, obtained through Eq.~\eqref{eq:HomogP} and the method of undetermined coefficients, is given by 
\begin{equation}
P=\me^{\gamma z} (C_1 \sinh\sigma z+C_2\cosh\sigma z)+\me^{M z}\left(\frac{b_1}{A-s}z+\frac{b_0}{A-s}-\frac{2b_1(M-\gamma)}{(A-s)^2} \right)s\Delta_y Q,
\label{eq:genpz2}
\end{equation}
where $A=M(M-2\gamma)$. The coefficients 
$C_1(s)$ and $C_2(s)$ are set in order to respect the two boundary conditions mentioned above, and then arise from a linear system of two equations, so that  
\begin{equation}
\begin{split}
\frac{P}{\Delta_y Q}&=\frac{\me^{\gamma z}}{\sinh\sigma}\left\{\left(\frac{Mb_o+b_1}{A-s}-\frac{2b_1M(M-\gamma)}{(A-s)^2}\right)\left[\cosh\sigma z(\gamma\sinh\sigma+\sigma\cosh\sigma-\me^{M-\gamma}\sigma)\right.\right.\\ &+\left.\left.
\sinh\sigma z(\gamma\me^{M-\gamma}-\gamma\cosh\sigma-\sigma\sinh\sigma)\right]+\frac{Mb_1}{A-s}\me^{M-\gamma}(\gamma\sinh\sigma z-\sigma\cosh\sigma z)\right\}\\ &+
\me^{M z}\left(\frac{b_1}{A-s}z+\frac{b_0}{A-s}-\frac{2b_1(M-\gamma)}{(A-s)^2} \right)s
.\end{split}
\label{eq:PQmz}
\end{equation}
At this point, only for the sake of brevity, we restrict our solution to the load application surface situated at $z=0$. We define $P_0=P|_{z=0}$ and obtain 
\begin{equation}
\begin{split}
\frac{P_0}{\Delta_y Q}&=\frac{1}{\sinh\sigma}\left\{\left(\frac{Mb_o+b_1}{A-s}-\frac{2b_1M(M-\gamma)}{(A-s)^2}\right)\left(\gamma\sinh\sigma+\sigma\cosh\sigma-\sigma\me^{M-\gamma}\right)\right.\\ &-\left.
\frac{Mb_1}{A-s}\sigma\me^{M-\gamma}\right\}+
\frac{sb_0}{A-s}-\frac{2b_1(M-\gamma)s}{(A-s)^2}
.\end{split}
\label{eq:P0Qz}
\end{equation}
The inverse Laplace transformation needs to be conducted on $P_0(s)$ in order to find $p_0(\tau)$. For this purpose we invert the ratio $\frac{P_0}{\Delta_y Q}$ and at the end, exploiting the convolution theorem, convolve the result with $\Delta_y q(\tau)$. While the inversion of the terms outside the curly brackets is trivial, the ones inside divided by $\sinh\sigma$ can be inverted making use of the residue theorem\footnote{The Laplace inversion of $f(s)$ can be done via contour integration as follows:\begin{equation}
\Laplinv{f(s)}=\dfrac{1}{2\pi\IM}\dinteg{c-\IM\infty}{c+\IM\infty}{f(s)\me^{s\tau}}{s}
\end{equation}
with $c\in\mathbb{R}$ greater than the real part of every singularity of $f(s)$. If the integral converges for $|s|\rightarrow\infty$ in the half-plane $\mathrm{Re}(s)<c$ and the singularities lie on the real axis, then it can be reduced to
\begin{equation}
\Laplinv{f(s)}=-\dfrac{1}{2\pi\IM}\dinteg{-\infty}{c}{f(s^{+})\me^{s^{+}\tau}}{s^{+}}-\dfrac{1}{2\pi\IM}\dinteg{-\infty}{c}{f(s^{-})\me^{s^{-}\tau}}{s^{-}}+\sum_{j=1}^N{\Res{f(s)\me^{s\tau}}{s_j}},
\end{equation}
in which $N$ is the number of the singularities located at $s_j$, $s^{+}$ means that the path of integration is taken above the real axis, $s^{-}$ vice versa. It is the case of $\Omega(s)$ (see Eq.~\eqref{eq:p20oms}), for which it is also easy to show the two integrals on the right-hand side eliminate each other, so that the inverse Laplace transform results simply the sum of the residues.}. The poles of such functions are in $s=A$ and where $\sinh\sigma$ vanishes. This occurs when $\sigma=n\pi\IM$, namely in $s=s_n=-(\gamma^2+n^2\pi^2)$ ($n=0,1,...,+\infty$), so that
\begin{equation}\begin{split}
\Laplinv{\frac{P_0}{\Delta_y Q}}&=\Res{\me^{s\tau}\Omega(s)}{s=A}+\sum_{n=0}^\infty{\Res{\me^{s\tau}\Omega(s)}{s=s_n}}\\&-b_0\left(A\me^{A\tau}+\delta(\tau)\right)-2b_1(M-\gamma)\me^{A\tau}(1+A\tau),
\end{split}\label{eq:P0Qzsol}
\end{equation}
where $\delta(t)$ is Dirac's delta function and
\begin{equation}
\begin{aligned}\Omega(s)&=\frac{1}{\sinh\sigma}\left\{\left(\frac{Mb_o+b_1}{A-s}-\frac{2b_1M(M-\gamma)}{(A-s)^2}\right)\left(\gamma\sinh\sigma+\sigma\cosh\sigma-\sigma\me^{M-\gamma}\right)\right.\\ &-\left.
\frac{Mb_1}{A-s}\sigma\me^{M-\gamma}\right\}.\end{aligned}
\label{eq:p20oms}
\end{equation}
It results that
\begin{equation}
\Res{\me^{s\tau}\Omega(s)}{s=A}=\me^{A\tau}Ab_0+2b_1(M-\gamma)\me^{A\tau}(1+A\tau).
\label{eq:P0QzRA}
\end{equation}
If $L_n$ is defined as $L_n=((M-\gamma)^2+n^2\pi^2)$, the residue in $s_n$ is written as
\begin{equation}\begin{split}
\Res{\me^{s\tau}\Omega(s)}{s=s_n}&=\left\{\left[\frac{Mb_o+b_1}{L_n}-\frac{2b_1M(M-\gamma)}{L_n^2}\right]\left((-1)^n\me^{M-\gamma}-1\right)\right.\\
&\left.+\frac{Mb_1}{L_n}(-1)^n\me^{M-\gamma}\right\}
2n^2\pi^2\me^{-(\gamma^2+n^2\pi^2)\tau},
\end{split}
\label{eq:P0QzRn}
\end{equation}
which vanishes at $s=0$ ($n=0$). Making use of the convolution theorem and collecting the expressions \eqref{eq:P0QzRA} and \eqref{eq:P0QzRn},  Eq.~\eqref{eq:P0Qzsol} gives
\begin{equation}
p_{0i}^{(M)}(\tau)=-b_{0i}\Delta_yq(\tau)+\sum_{n=1}^\infty{\mathrm{Res}\left\{\me^{s\tau}\Omega_i^{(M)}(s);s_{n}\right\}}\ast\Delta_yq(\tau),
\end{equation}
where the index $i$ and the superscript $(M)$ have the same meaning as in Eq.~\eqref{eq:LMn2+3}.
Now we proceed to evaluate $\Xi_0=\Laplinv{\xi}|_{z=0}=-\dinteg{0}{1}{P\me^{-2\gamma z}}{z}$, where $P$ is defined in Eq.~\eqref{eq:PQmz}. In this way, we obtain 
\begin{equation}\begin{split}
\frac{\Xi_0}{\Delta_yQ}&=-\me^{M-2\gamma}\left(\frac{Mb_0+(M+1)b_1}{s-A}+\frac{2b_1M(M-\gamma)}{(s-A)^2}\right)\\
&+\left(\frac{Mb_0+b_1}{s-A}+\frac{2b_1M(M-\gamma)}{(s-A)^2}\right)+\frac{s}{s-A}\frac{b_1}{M-2\gamma}\me^{M-2\gamma}\\
&+\frac{\me^{M-2\gamma}-1}{M-2\gamma}\left[\frac{s}{s-A}b_0+\frac{2sb_1(M-\gamma)}{(s-A)^2}-\frac{sb_1}{(M-2\gamma)(s-A)}\right]
.\end{split}
\label{eq:Xi0QM}
\end{equation}
Passing to inverse Laplace transforms and convolving by $\Delta_yq(\tau)$, each $i$-th term of $\xi_0^{(M)}=\sum_{i=1}^4\xi_{0i}^{(M)}$ results to be
\begin{equation}
\begin{split}
\xi_{0i}^{(M)}(\tau)&=\left[b_{0i}\frac{\me^{M_i-2\gamma}-1}{M_i-2\gamma}+b_{1i}\frac{\me^{M_i-2\gamma}(M_i-2\gamma-1)+1}{(M_i-2\gamma)^2}\right]\Delta_yq(\tau)\\
&+b_{1i}M_i(\me^{M_i-2\gamma}-1)\me^{A_i\tau}\ast\Delta_yq(\tau)
.\end{split}
\label{eq:xi0QMt}
\end{equation}
\section{Solution to Eq.~\eqref{eq:LMn2+3} for the right-hand sides $\Upsilon_5(s,z)$}
\label{1otAppendix_B}
The equation, coupled with the usual boundary conditions on the derivative of P, is
\begin{equation}
\ppder{P}{z}-2\gamma\pder{P}{z}-s P=
-\frac{z^2}{2}s\Delta_y Q,
\label{eq:Mp2}
\end{equation}
and its solution is given by 
\begin{equation}
P=\me^{\gamma z} (C_1(s) \sinh\sigma z+C_2(s)\cosh\sigma z)+\left(\frac{z^2}{2}-\frac{2\gamma z}{s}+\frac{1}{s}+\frac{4\gamma^2}{s^2} \right)\Delta_y Q,
\label{eq:Pz2comp}
\end{equation}
where
\begin{gather}
C_1(s)=\frac{\frac{2\gamma}{s}(\gamma\cosh\sigma+\sigma\sinh\sigma-\gamma\me^{-\gamma})+ \gamma\me^{-\gamma}        }{s \sinh\sigma}\Delta_y Q,\\
C_2(s)=\frac{\frac{2\gamma}{s}(\sigma\me^{-\gamma}-\gamma\sinh\sigma-\sigma\cosh\sigma)-\sigma\me^{-\gamma}        }{s \sinh\sigma}\Delta_y Q.
\label{eq:c1c2pz2}
\end{gather}
Complying exclusively with the search for $P_0=P|_{z=0}$, the next function must be inverted to give 
\begin{equation}\begin{split}
\frac{P_0}{\Delta_y Q}&=\frac{\frac{2\gamma}{s}(\sigma\me^{-\gamma}-\gamma\sinh\sigma-\sigma\cosh\sigma)-\sigma\me^{-\gamma}        }{s \sinh\sigma}+\frac{1}{s}+\frac{4\gamma^2}{s^2}\\
&=\Omega(s)+\frac{1}{s}+\frac{4\gamma^2}{s^2},
\end{split}
\end{equation} 
with the self-evident definition of $\Omega(s)$. The latter has non-zero residues at $s=0$ and at $s=s_n$. Similarly to what explained in the previous paragraph, the inverse Lalace transform of $\Omega(s)$ can be executed only in terms of residues
\begin{equation}\begin{split}
\Laplinv{\frac{P_0}{\Delta_y Q}}&=\Res{\Omega(s)\me^{s\tau}}{0}+\sum_{n=0}^{\infty}\Res{\Omega(s)}{s_n}+1+4\gamma^2\tau.
\end{split}
\label{eq:P0z2}
\end{equation} 
On the other hand, it occurs that $\Res{\Omega(s)}{0}=-1-4\gamma^2t$. It remains to compute the residue at $s=s_n$:
\begin{equation}
\Res{\me^{s\tau}\Omega(s)}{s_n}=\left[2\gamma\frac{1-(-1)^n\me^{-\gamma}}{(\gamma^2+n^2\pi^2)^2}-\frac{(-1)^n\me^{-\gamma}}{\gamma^2+n^2\pi^2}\right]2n^2\pi^2\me^{-(\gamma^2+n^2\pi^2)\tau}.
\end{equation}
Substituting this results into Eq.~\eqref{eq:P0z2} and taking advantage from the convolution theorem, it follows that
\begin{equation}
p_0^{(2)}(\tau)=\sum_{n=1}^{\infty}\Res{\Omega^{(2)}(s)\me^{s\tau}}{s_n}\ast\Delta_y q(\tau),
\end{equation}
where we use the superscript $(2)$ to signify $z^2$-term. With respect to $\Xi_0$, the integration of Eq.~\eqref{eq:Pz2comp} along the layer depth yields
\begin{equation}\begin{split}
\frac{\Xi_0}{\Delta_yQ}&=-\frac{C_1(s)}{\Delta_yQ}\dinteg{0}{1}{\me^{-\gamma z}\sinh\sigma z}{z}-\frac{C_2(s)}{\Delta_yQ}\dinteg{0}{1}{\me^{-\gamma z}\cosh\sigma z}{z}\\
&-\dinteg{0}{1}{\left(\frac{z^2}{2}-\frac{2\gamma z}{s}+\frac{1}{s}+\frac{4\gamma^2}{s^2} \right)}{z}.\end{split}
\end{equation}
The latter, after substituting the costants of Eq.~\eqref{eq:c1c2pz2} leads simply to
\begin{equation}
\xi_0^{(2)}(\tau)=-\Delta_yq(\tau)\dinteg{0}{1}{\me^{-2\gamma z}\frac{z^2}{2} }{z}
.\label{eq:xi02t}
\end{equation}
\section{Solution to Eq.~\eqref{eq:LMn2+3} for the right-hand sides $\Upsilon_5(s,z)$}
\label{1otAppendix_C}
The function $P$, regarding the $k$-term in Eq.~\eqref{eq:LMn2+3}, can be directly recovered from the solution of Eq.~\eqref{eq:genpz2}, setting $b_1$ to zero, $b_0=-\frac{1}{s}\frac{k_1}{k_3}$, $M=2\gamma-\gamma_1$ and $A=-\gamma_1(2\gamma-\gamma_1)$. Thus, we will have 
\begin{equation}\begin{split}
\frac{P}{\Delta_yQ}&=\frac{\me^{\gamma z}Mb_0}{(A-s)\sinh\sigma}\left\{\cosh\sigma z(\gamma\sinh\sigma+\sigma\cosh\sigma-\sigma\me^{M-\gamma})\right.\\&+\left.
\sinh\sigma z(\gamma\me^{M-\gamma}-\gamma\cosh\sigma-\sigma\sinh\sigma)\right\}
+\me^{\gamma z}\frac{b_0}{A-s}
.\end{split}
\end{equation}
On the load application surface, we have 
\begin{equation}\begin{split}
\frac{P_0}{\Delta_yQ}&=\frac{M\frac{k_1}{k_3}}{s(s-A)\sinh\sigma}(\gamma\sinh\sigma+\sigma\cosh\sigma-\gamma\me^{M-\gamma})+\frac{k_1}{k_3}\frac{1}{s-A}\\
&=\Omega(s)+\frac{k_1}{k_3}\frac{1}{s-A},
\end{split}
\end{equation}
where, again, the meaning of $\Omega(s)$ is clear. The inverse Laplace transform of $\Omega(s)$ is calculated via residue theorem as the sum of the residues of $\Omega(s)\me^{st}$ at the singularities, which are situated at $0$, $A$ and $s_n$.
\begin{gather}
\Res{\Omega(s)\me^{st}}{0}=\frac{\gamma \frac{k_1}{k_3}}{(M-2\gamma)\sinh\gamma}(\me^{M-\gamma}-\me^\gamma),\\
\Res{\Omega(s)\me^{s\tau}}{A}=-\frac{k_1}{k_3}\me^{A\tau},\\
\Res{\Omega(s)\me^{s\tau}}{s_n}=\frac{M(1-(-1)^n\me^{M-\gamma})}{\left((M-\gamma)^2+n^2\pi^2\right)(\gamma^2+n^2\pi^2)}2n^2\pi^2\me^{-(\gamma^2+n^2\pi^2)\tau}.
\end{gather}
Connoting with $(k)$ the elements arising from the $k$-terms, the complete inverse Laplace transform gives
\begin{equation}
p_0(\tau)=\frac{\gamma \frac{k_1}{k_3}\me^\gamma}{\gamma_1\sinh\gamma}(1-\me^{-\gamma_1})\ast\Delta_yq(\tau)+\sum_{n=1}^\infty{\Res{\me^{s\tau}\Omega^{(k)}(s)}{s_n}}\ast\Delta_yq(\tau).
\end{equation}
The function $\Xi_0(s)$ is (using the same substitution as explained for the pressure) simply recovered from Eq.~\eqref{eq:Xi0QM} and is
\begin{equation}
\frac{\Xi_0(s)}{\Delta_yQ}=\frac{k_1}{k_3}\left(\me^{-\gamma_1}-1\right)\left[\frac{2\gamma-\gamma_1}{s(s-A)}+\frac{1}{\gamma_1(s-A)}   \right].
\end{equation}
Inverting the latter Laplace transform and using the convolution theorem, finally we obtain
\begin{equation}
\xi_0^{(k)}(\tau)=\frac{k_1}{k_3}\frac{\me^{-\gamma_1}-1}{\gamma_1}\ast\Delta_yq(\tau).
\label{eq:xi0kt}
\end{equation}

\end{document}